\documentclass[fleqn,twoside]{article}
\usepackage{espcrc2}

\usepackage{graphicx}
\usepackage[figuresright]{rotating}

\newcommand{\pom}{{I\!\!P}}
\newcommand{\reg}{{I\!\!R}}
\newcommand{\xpom}{\ensuremath{{x_\pom}}}
\newcommand{\fpom}{\ensuremath{f_{\pom}}}
\newcommand{\freg}{\ensuremath{f_{\reg}}}
\newcommand{\apom}{\ensuremath{\alpha_\pom}}
\newcommand{\appom}{\ensuremath{\alpha'_\pom}}
\newcommand{\ftd}{\ensuremath{{F_2^D}}}
\newcommand{\fld}{\ensuremath{{F_L^D}}}
\newcommand{\srd}{\ensuremath{{\sigma_r^D}}}
\renewcommand\figurename{Fig.}
\newcommand\figuresname{Figs.}
\newcommand{\eqref}[1]{Eq.~(\ref{#1})}
\newcommand{\figref}[1]{\figurename~\ref{#1}}
\newcommand{\figsref}[2]{\figuresname~\ref{#1} and~\ref{#2}}
\newcommand{\figrange}[2]{\figuresname~\ref{#1}--\ref{#2}}
\newcommand{\secref}[1]{Sec.~\ref{#1}}
\def\Journal#1#2#3#4{{#1} {\bf #2} (#3) #4}

\def\NPB{{\em Nucl. Phys.}   {\bf B}}

\def\PRD{{\em Phys. Rev.}    {\bf D}}

\def\SJNP{{\em Sov. J. Nucl. Phys.}}
\def\SPJ{{\em Sov. Phys. JETP}}

\title{H1 Diffractive Structure Function Measurements and QCD Fits}

\author{S.~Sch\"atzel\address{Physikalisches Institut,
        Philosophenweg 12, D--69120 Heidelberg, Germany}\thanks{Talk
        presented at Diffraction 2004 Conference, Sardinia.} (for the H1 Collaboration)}       
\begin{document}

\begin{abstract}
Measurements of diffractive structure functions in $ep$ collisions
and diffractive parton densities extracted from QCD fits are presented.
\vspace{1pc}
\end{abstract}

\maketitle

\section{DIFFRACTION AT HERA}
\begin{figure}[hhh] 
\vspace{-0.5cm}
\centering
\includegraphics[width=0.35\textwidth,keepaspectratio]{%
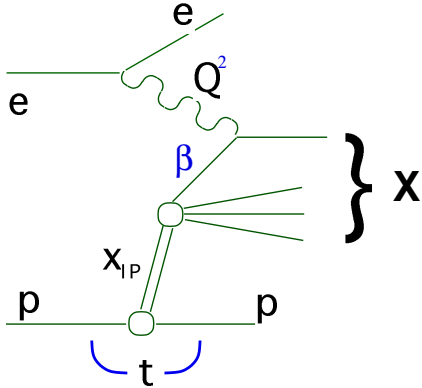}
\vspace{-1cm}
\caption{Diffractive $ep$ scattering.}
\label{fig1}
\vspace{-0.5cm}
\end{figure}

At the HERA $ep$ collider the diffractive quark structure of the proton
is probed with a point-like photon (\figref{fig1}). 
The virtuality of the photon is denoted by $Q^2$ and sets the hard
scale of the interaction. 
Diffraction is characterised by an elastically scattered proton
which loses only a small fraction $\xpom$ of its initial beam
momentum.
These events are selected experimentally by detecting the proton
at small scattering angles (roman pot detectors)
or by requiring a large empty area in the detector between the outgoing
proton and the hadronic system $X$ produced in the interaction (rapidity gap).
The squared 4-momentum $t$ transferred at the proton vertex can be 
measured by tagging the proton.
For the rapidity gap method, which accesses a much larger event
sample, the cross section has to be integrated over $|t|<1$~GeV$^2$ and in
$\approx 10\%$ of the events the proton is excited into a hadronic
system of small mass $<1.6$~GeV. The two methods give the same
results when compared in the same kinematic range.

In a picture which depicts diffraction as a two step process,
the proton exchanges a diffractive object (often called
the pomeron) with momentum fraction $\xpom$ and the quark struck by the 
photon carries a fraction $\beta$ of the momentum of the diffractive
exchange.
Additional kinematic variables are Bjorken-x $x=\beta \xpom$ and
the inelasticity $y=Q^2/(s x)$ where $s$ is the $ep$ centre-of-mass
energy squared.

The cross section is proportional to the combination of two structure
functions:
\begin{equation}
\sigma \propto \ftd - Y \fld \equiv \srd, \label{eq_ftd}
\end{equation}
where $Y=\frac{y^2}{1+(1-y)^2}$ is a kinematic factor 
 resulting from the difference of
the fluxes of transversely and longitudinally polarised photons from
the electron. 
$\ftd$ is proportional to the diffractive $\gamma^* p$ cross section,
whereas $\fld$ is related only to the part induced by longitudinal photons.
The factor $Y$ is sizable at large values of $y$ and
in most of the phase space measured so far, $\fld$ is a small
correction.

\section{FACTORISATION IN DIFFRACTION}
The diffractive structure functions have been proven to factorise
into diffractive parton densities $f_i^D$ of the proton convoluted
with ordinary photon-parton scattering cross sections 
$\hat{\sigma}^{\gamma^* i}$~\cite{Collins}:
\begin{equation}
\ftd = \sum_i f_i^D \otimes \hat{\sigma}^{\gamma^* i},
\end{equation}
where the sum runs over all partons. This factorisation formula holds
for large enough scales at leading twist and applies also to $\fld$.
The diffractive parton densities obey the standard QCD evolution
equations and can be determined from fits to structure function data.

\section{EXPERIMENTAL RESULTS}
\subsection{Dependence on \boldmath{$t$}}
\begin{figure}[hhh] 
\vspace{-0.5cm}
\centering
\includegraphics[width=0.45\textwidth,keepaspectratio]{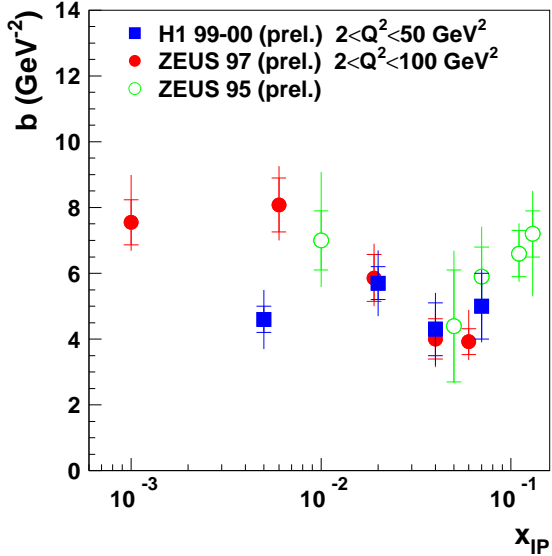}
\vspace{-1cm}
\caption{The slope parameter $b$ from fits to the diffractive $ep$ 
cross section $d\sigma/dt \propto e^{b t}$ for different values
of \xpom.}
\label{figt}
\vspace{-0.5cm}
\end{figure}

The $t$ dependence of the cross section has the form 
$d\sigma/dt \propto e^{b t}$ with a 
slope parameter $b\approx$ 5~to~7~GeV$^{-2}$ 
which at the present level of precision does
not depend on $\xpom$ as shown in \figref{figt}~\cite{fps}.

\subsection{Dependence on \boldmath{$\xpom$}}
\label{sec_xpom}
\begin{figure}[hhh]
\vspace{-0.5cm}
\centering
\includegraphics[width=0.48\textwidth,keepaspectratio]{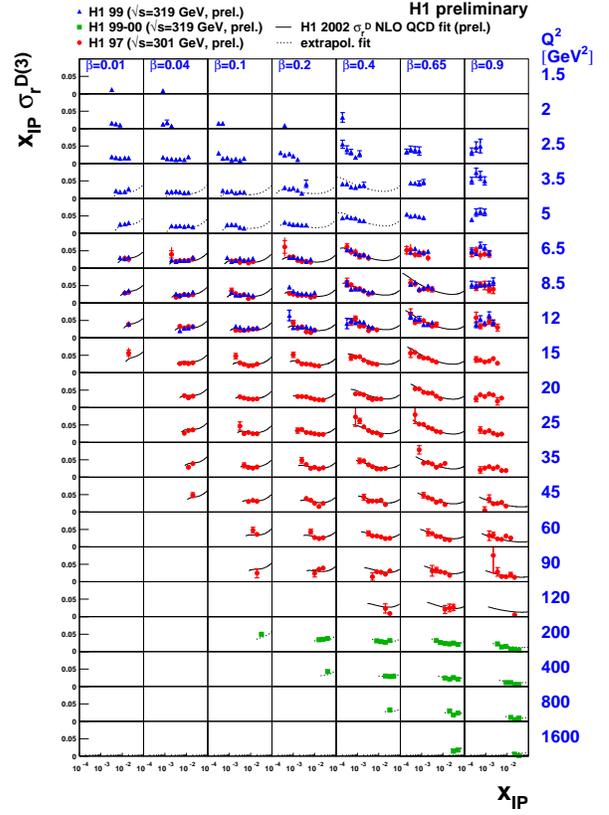}
\vspace{-1cm}
\caption{The reduced diffractive $ep$ cross section $\srd \equiv \ftd
  - Y \fld$ compared with the H1 2002 NLO QCD fit.}
\label{figsrd}
\vspace{-0.5cm}
\end{figure}

\begin{figure}[hhh]
\vspace{-0.5cm}
\centering
\includegraphics[width=0.48\textwidth,keepaspectratio]{%
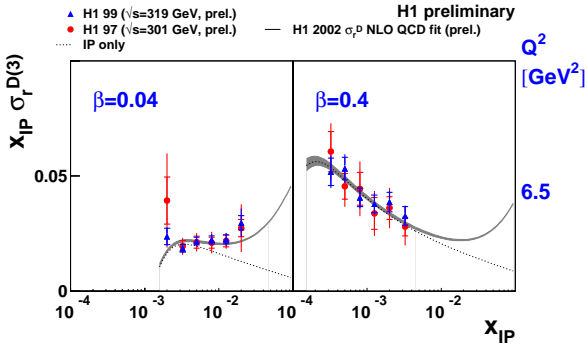}
\vspace{-1cm}
\caption{The $\xpom$ dependence of the reduced diffractive cross
  section for low and high $\beta$ compared with
the H1 2002 NLO QCD fit.}
\label{figmeson}
\vspace{-0.5cm}
\end{figure}

The reduced diffractive cross section is shown in \figref{figsrd}
as a function of \xpom{} in bins of $\beta$ and $Q^2$.
The measurements are obtained using a rapidity gap selection and 
cover a large kinematic range $Q^2=1.5$~to~1600~GeV$^2$~\cite{fit,gap}.
The collected event sample statistics do not allow an extraction of diffractive
parton densities at fixed values of $\xpom$.
Instead, the $\xpom$ and $t$ dependence of the PDFs are parameterised 
in a so-called flux factor \fpom:
\begin{equation}
f_i^D(\beta, Q^2, \xpom, t) = f_{\pom}(\xpom,t) \ f^D_i(\beta, Q^2) 
\end{equation}
with $\fpom(\xpom,t) = e^{b t} \xpom^{1-2 \apom(t)}$, 
where $\apom(t) = \apom(0) + \appom\,t$ is the linear
pomeron Regge trajectory.
This flux factor approach is consistent with the data within
the present uncertainties for $\xpom<0.01$. 
At larger $\xpom$ values, a second term has to be introduced which can be
interpreted as reggeon exchange:
\begin{equation}
\srd =  \fpom \left( F_2^D + Y F_L^D \right)
      + \freg  \left( F_2^\reg + Y F_L^\reg \right).
\end{equation}
This is illustrated in \figref{figmeson} where the 
cross section is well described by \fpom{} alone for
$\xpom<0.01$.
A fit to the data gives an
intercept $\alpha_\pom(0) = 1.17^{+0.07}_{-0.05}$ which
is larger than 1.08 as obtained for the soft pomeron in 
hadron-hadron collisions.

\subsection{Scaling Properties and \boldmath{$\beta$} dependence}
\begin{figure}[hhh]
\vspace{-0.5cm}
\centering
\includegraphics[width=0.48\textwidth,keepaspectratio]{%
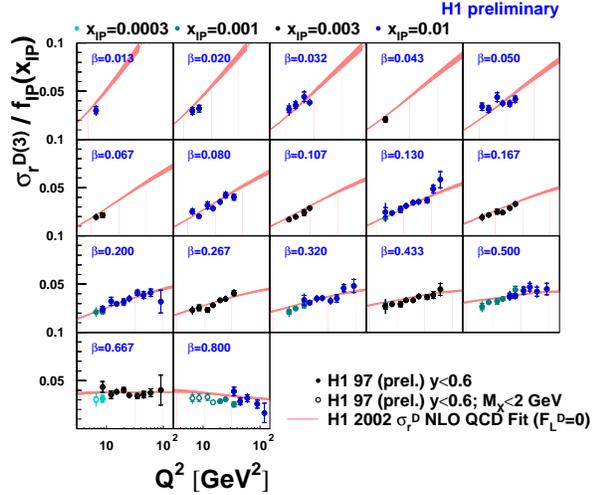}
\vspace{-1cm}
\caption{The diffractive structure function $\ftd$ as a function of
  $Q^2$ compared with the H1 2002 NLO QCD fit. The $\xpom$ dependence
  has been divided out.}
\label{figqs}
\vspace{-0.5cm}
\end{figure}

\begin{figure}[hhh]
\vspace{-0.5cm}
\centering
\includegraphics[width=0.48\textwidth,keepaspectratio]{%
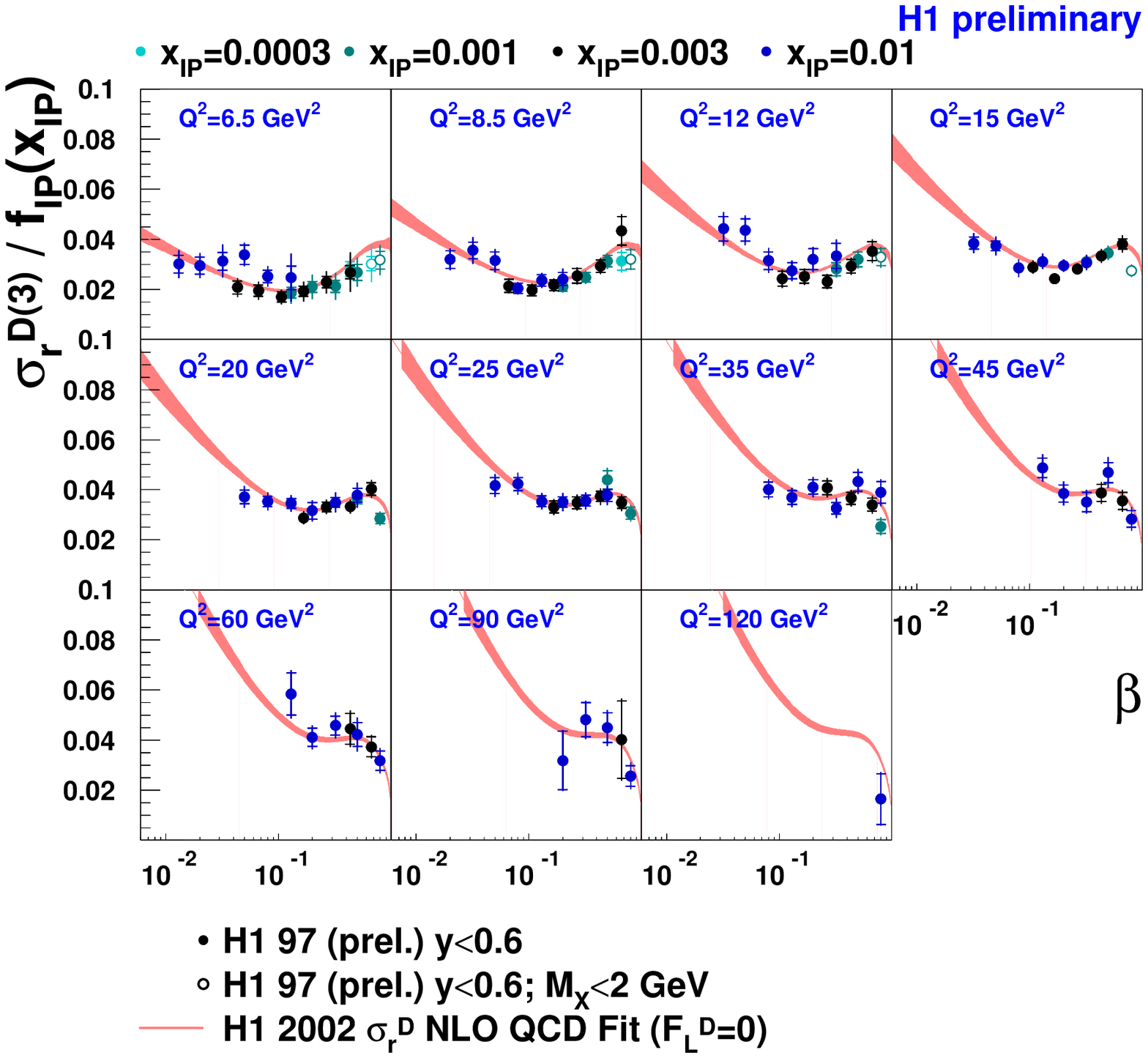}
\vspace{-1cm}
\caption{The diffractive structure function $\ftd$ as a function of
  $\beta$ compared with the H1 2002 NLO QCD fit. The $\xpom$ dependence
  has been divided out.}
\label{figbeta}
\vspace{-0.5cm}
\end{figure}

The $Q^2$ and $\beta$ dependences of the diffractive cross section are shown 
in \figsref{figqs}{figbeta} in a kinematic range ($y<0.6$, $\xpom<0.01$)
where to good approximation $\srd = \ftd$ and the reggeon term is
negligible. 
The shown data points display the pure $\beta$ and $Q^2$ dependences
of the structure function $\ftd$;
kinematic effects related to $\xpom$ and $t$ have been
corrected by dividing the cross section by $\fpom$.
In \figref{figqs} the structure function displays approximate scaling
for $\beta=2/3$. For lower values the data exhibit scaling violations
which are driven by a large gluonic component in the diffractive exchange.
The structure function depends only weakly on $\beta$ as shown
in \figref{figbeta}.

\subsection{Diffractive parton densities}
\begin{figure}[hhh]
\vspace{-0.5cm}
\centering
\includegraphics[width=0.48\textwidth,keepaspectratio]{%
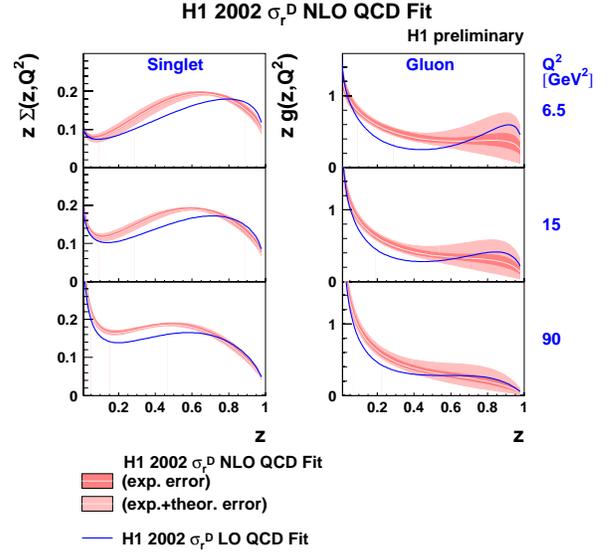}
\vspace{-1cm}
\caption{The H1 diffractive quark and gluon densities as extracted in a
  QCD fit to structure function data.}
\label{figpdf}
\vspace{-0.5cm}
\end{figure}

The H1 Collaboration has extracted diffractive parton densities
from QCD fits to the diffractive structure function data.
The $\xpom$ and $t$ dependence of the PDFs is given by the flux factor
as discussed in \secref{sec_xpom}. The $\beta$ dependences of the quark
and gluon densities are
parameterised at a starting scale $Q_0^2=3~$GeV$^2$ and are evolved
to the measured $Q^2$ values using the DGLAP equations~\cite{dglap}.
The best fit is shown in \figrange{figsrd}{figbeta} and
describes the measurements very well.
The corresponding parameterisations for the NLO and LO quark and gluon densities are shown in
\figref{figpdf}.
The gluon carries $\approx 75\%$ of the momentum of the diffractive
exchange.
The error band around the NLO densities includes 
experimental (inner band) and model uncertainties (outer band) 
which have been propagated to the PDFs.
The gluon density is known to better than 30\% up to fractional
momenta $z \approx 0.5$, but is poorly known at large $z$.
These densities have been used to predict diffractive final state cross
sections such as dijet and heavy flavour production at HERA~\cite{svetlana} and at the
Tevatron~\cite{fit}.

\subsection{Ratio of diffractive to inclusive cross section}
\begin{figure}[hhh]
\vspace{-0.5cm}
\centering
\includegraphics[width=0.35\textwidth,keepaspectratio]{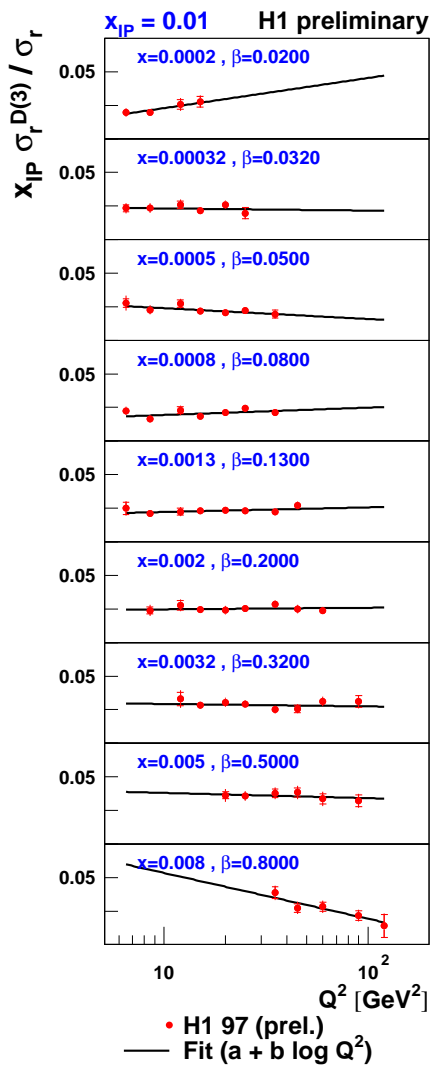}
\vspace{-1cm}
\caption{The ratio of diffractive to inclusive $ep$ scattering cross
  sections as a function of $Q^2$ at $\xpom=0.01$ for different values of $x=\beta \xpom$.}
\label{figratio}
\vspace{-0.5cm}
\end{figure}

The ratio of the diffractive to the inclusive cross section at the
same $x=\beta \xpom$ is shown
in \figref{figratio} for $\xpom=0.01$ as a function of $Q^2$.
For this particular $\xpom$ and the corresponding gap size, 
the diffractive contribution amounts to
2--3\% of the inclusive cross section.
The ratio 
is flat for $\beta<0.6$ indicating a similar QCD evolution of
the inclusive and the diffractive structure functions away from the
kinematic limit $\beta=1$~\cite{fit}.


\subsection{Charged current cross section}
\begin{figure}[hhh]
\vspace{-0.5cm}
\centering
\includegraphics[width=0.49\textwidth,keepaspectratio]{%
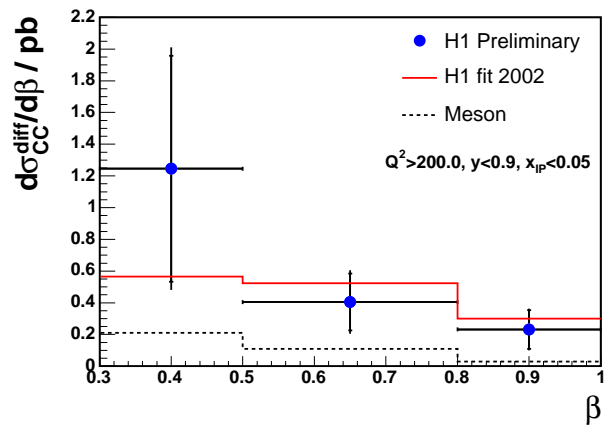}
\vspace{-1cm}
\caption{The diffractive charged current cross section as a function
  of the momentum fraction $\beta$ compared with a LO prediction based
  on the LO diffractive parton densities of \figref{figpdf}.}
\label{figcc}
\vspace{-0.5cm}
\end{figure}

Diffractive processes which occur via $W$ boson exchange instead of
photon exchange have been measured by H1 using events with missing transverse energy which is
carried away by the neutrino~\cite{cc}.
The ratio of the diffractive to the inclusive charged current cross section was measured
to be $2.5\% \pm 1.0\%$ for $\xpom<0.05$.
The cross section as a function of $\beta$ is shown in \figref{figcc}.
It is well described by a leading order Monte Carlo prediction which
is based on the diffractive parton densities extracted in neutral
current processes.

\section{CONCLUSIONS}
Diffractive structure functions have been measured by the H1
Collaboration to unprecedented precision. 
The data are consistent with QCD factorisation and
diffractive parton densities have been extracted in QCD evolution
fits. The gluon component 
carries $\approx 75\%$ of the momentum of the diffractive exchange.
QCD factorisation was tested in diffractive charged current
interactions where 
predictions based on the neutral current PDFs are in good agreement
with the measured cross section.
Diffractive and inclusive deep-inelastic $ep$ scattering
were shown to evolve similarly with the hard QCD scale.

\subsection*{Acknowledgements}
I thank my colleagues in H1 for their work reflected in this article
and the organisers for an interesting conference in a
splendid location.
I thank Nicholas Malden for a thorough reading of the manuscript.

\end{document}